\documentclass[12pt]{article}
\usepackage{amsmath}
\usepackage{enumerate}
\usepackage{natbib}
\usepackage{url} 
\usepackage{graphicx}
\RequirePackage{amsthm,amsmath,amsfonts,amssymb}
\usepackage{hyperref}
\RequirePackage{bm}
\usepackage{natbib}
\usepackage{caption}
\usepackage{multirow}
\usepackage{multicol}
\usepackage{enumerate}
\usepackage{cprotect}
\usepackage{float}
\usepackage[caption=false]{subfig}
\usepackage{booktabs}
\usepackage[table,xcdraw]{xcolor}

\newcommand{\blind}{0}

\theoremstyle{plain}

\newtheorem{theorem}{Theorem}

\begin{document}

\def\spacingset#1{\renewcommand{\baselinestretch}%
{#1}\small\normalsize} \spacingset{1}


\if0\blind
{
  \title{\bf A novel longitudinal rank-sum test for multiple primary endpoints in clinical trials: Applications to neurodegenerative disorders}
  \author{Xiaoming Xu, Dhrubajyoti Ghosh and Sheng Luo$^\ast$ \\
    Department of Biostatistics and Bioinformatics\\
    Duke University, Durham, NC, USA}
  \maketitle
} \fi

\if1\blind
{
  \bigskip
  \bigskip
  \bigskip
  \begin{center}
    {\LARGE\bf Title}
\end{center}
  \medskip
} \fi

\bigskip
\begin{abstract}
Neurodegenerative disorders such as Alzheimer's disease (AD) present a significant global health challenge, characterized by cognitive decline, functional impairment, and other debilitating effects. Current AD clinical trials often assess multiple longitudinal primary endpoints to comprehensively evaluate treatment efficacy. Traditional methods, however, may fail to capture global treatment effects, require larger sample sizes due to multiplicity adjustments, and may not fully exploit multivariate longitudinal data. To address these limitations, we introduce the Longitudinal Rank Sum Test (LRST), a novel nonparametric rank-based omnibus test statistic. The LRST enables a comprehensive assessment of treatment efficacy across multiple endpoints and time points without multiplicity adjustments, effectively controlling Type I error while enhancing statistical power. It offers flexibility against various data distributions encountered in AD research and maximizes the utilization of longitudinal data. Extensive simulations and real-data applications demonstrate the LRST's performance, underscoring its potential as a valuable tool in AD clinical trials.
\end{abstract}

\noindent%
\vfill

\newpage
\spacingset{1.45} 

\section{Introduction}
Alzheimer's disease (AD), a prevalent neurodegenerative disorder, affects millions globally, with a substantial 6.7 million cases in the United States alone \citep{2023ADJ_facts_figures}. It is characterized by progressive cognitive and functional deterioration, significantly impairing patient autonomy and caregiver quality of life. The etiology of AD involves a complex interplay of genetic, environmental, and lifestyle factors, leading to significant individual variability in its manifestation and progression \citep{Scheiblich2020Trends-Immunology}. This heterogeneity poses considerable challenges in developing efficacious treatments. In response, the US Food and Drug Administration (FDA) recommends incorporating both cognitive and functional outcomes as primary endpoints in AD clinical trials, in 2018 guidelines \citep{FDA2018AD_Guidance}. Current AD randomized clinical trials (RCTs) often capture multiple longitudinal primary endpoints, encompassing neuropsychological, functional, and behavioral assessments, and increasingly, blood-based biomarkers \citep{Hansson2023Nature-Aging}. An analysis of 314 late-phase AD efficacy trials conducted since 1996 shows that nearly half (155 trials or 49.4\%) included two or more primary endpoints, reflecting a shift towards a more comprehensive evaluation paradigm \citep{Yu2022ADJ-TRCI}.

In evaluating multivariate longitudinal outcomes from AD RCTs that compare experimental treatments against placebos or active controls, two primary methodological approaches are prevalent. The first, the individual test approach, assesses longitudinal outcomes either independently or jointly to determine treatment effects for each outcome, followed by multiplicity adjustment. Analytical techniques include parametric methods such as univariate or multivariate GEE and linear mixed models (LMMs), as well as nonparametric methods such as univariate or multivariate functional mixed models (FMMs) \citep{Zhu2019Stat-Sinica} and various rank-based tests \citep{Akritas1997JSPI, Konietschke2010CSDA, Brunner2017JRSSB, Zhuang2018SMMR, Umlauft2019JMVA, Rubarth2022SMMR, Rubarth2022Symmetry}. Multiplicity adjustment may employ co-primary endpoint strategy, requiring significance across all endpoints for treatment efficacy assertion, adopted in studies such as Semagacestat \citep{Doody2013NEJM_Semagacestat}, EXPEDITION \citep{Doody2014NEJM}, LMTM \citep{Gauthier2016Lancet}, and ADMET 2 \citep{Mintzer2021JAMA-Neurology}, or multiple endpoint strategy, requiring at least one significant outcome after multiple comparison corrections (e.g., Bonferroni, Holm, or Hochberg procedures, and fixed-sequence methods) per FDA guidelines \citep{FDA_multiendpt_GUIDANCE2022}. However, these methods provide treatment effect estimates for individual outcomes, lacking a comprehensive assessment of global treatment efficacy across multiple endpoints when considering various disease aspects simultaneously. Additionally, Multiplicity adjustments can reduce power, requiring larger sample sizes \citep{Hamasaki2018JBS}. Moreover, LMMs' treatment efficacy estimation may be biased by endpoint non-normality arising from skewness, outliers, and their combination \citep{Lachos2011Biometrics}.

The second methodology in analyzing multivariate longitudinal outcomes in AD RCTs involves transforming longitudinal data into a cross-sectional format, focusing on changes from baseline to the final observation. This approach typically employs global testing procedures, which are broadly categorized into parametric and nonparametric methods. Parametric procedures, reliant on multivariate normality, include likelihood ratio tests \citep{Bartholomew1961JRSSB,Perlman1969AMS,Tang1994JASA} and $t$-statistics type tests \citep{Tang1993Biometrics,Bittman2009Biometrika,Lu2013CSDA}. However, their practical use is often limited due to stringent distributional assumptions. In contrast, nonparametric rank-based procedures, which are distribution-free, are widely used in clinical research settings, including multivariate factorial designs \citep{Munzel2000JSPI,Gunawardana2019JMVA}, multivariate categorical data \citep{Brunner2002JSPI}, cluster-correlated data \citep{Larocque2010JNS}, subjects with dependent replicates \citep{Roy2019SIM}, multivariate crossed layouts \citep{Dobler2020AISM}, and high-dimensional genetic data \citep{Meng2022JSSC}. Specifically, \cite{Brunner2002JSPI} developed Wald-type and ANOVA-type tests for simultaneous multiple outcome assessments. \cite{Roy2019SIM} addressed nonparametric methods for two-sample problems with correlated replicates. \cite{OBrien1984Biometrics} proposed a rank-sum-type test, which was later refined by \cite{Huang2005Biometrics} through a variance adjustment to control Type I error rate. \cite{Liu2010JASA} introduced a rank-max-type test, effective in detecting group differences across all outcomes, particularly when these differences are directionally opposite. Additionally, \cite{Zhang2019Biometrics} developed a cluster-adjusted rank-based test, combining the strengths of both rank-sum and rank-max tests by aggregating within-cluster evidence and combining it at the cluster level. Despite these advancements, these rank-based tests are primarily applied to cross-sectional data and cannot fully utilize multivariate longitudinal data. Consequently, they are inadequate for tracking temporal variations in treatment efficacy, which is crucial for a comprehensive assessment of treatment effects over the entire treatment period. 

Addressing these gaps, this article introduces the Longitudinal Rank Sum Test (LRST), a novel nonparametric rank-based omnibus test designed to comprehensively evaluate treatment efficacy across multiple longitudinal primary endpoints in AD RCTs. The LRST offers a robust and efficient statistical framework for detecting treatment effects across multiple endpoints over time. Compared to existing methods, the LRST innovates in several key areas. Firstly, it enables a global assessment of treatment efficacy across multiple endpoints and time points without requiring multiplicity adjustments. This approach enhances understanding of treatment effects and provides a holistic perspective on interventions in AD trials. Secondly, the LRST effectively controls Type I error while enhancing statistical power, thereby requiring smaller sample sizes. This efficiency reduces subjects' exposure to potentially ineffective treatments and lowers the financial burden of therapy development in AD RCTs. Thirdly, the LRST employs a nonparametric approach based on ranking, akin to the Wilcoxon rank-sum test, without relying on distributional assumptions such as normality in LMMs. This flexibility ensures robustness against various data distributions commonly encountered in AD research. Lastly, the LRST procedure maximizes the utilization of all longitudinal data, capturing dynamic changes across multiple endpoints and providing a comprehensive analysis of treatment effects in AD RCTs.

The rest of the article is organized as follows. Section~\ref{sec:Methods} develops LRST procedure and examines its theoretical properties. In Section~\ref{sec:Simulations}, extensive simulations are conducted to investigate the performance in Type I error and power. In Section~\ref{sec:RealData}, we applied the LRST procedure to two RCTs of neurodegenerative disorder: the Bapineuzumab (Bapi) 301 AD study and the Azilect Parkinson's disease (PD) study. Some discussions and future work are given in Section~\ref{sec:Discussions}. All technical details and additional results are provided in the Supplementary material.

\section{Statistical Framework} \label{sec:Methods}
\subsection{Notation and Asymptotic Distribution} \label{sec:Methods_theory}
In AD RCTs with multiple longitudinal outcomes and two parallel arms, treatment vs. control, the data structure is often of the type ($x_{itk}, y_{jtk}, t$), where $x_{itk}$ is the change of outcome $k$ ($k=1, \ldots K$ for a total of $K$ outcomes) from baseline to time $t$ ($t=1, \ldots T$, where $T$ is the number of after-baseline visits) from subject $i$ ($i=1, \ldots n_x$) in the control group, while $y_{jtk}$ is defined similarly for subject $j$ ($j=1, \ldots n_y$) in the treatment group, $n_x$ and $n_y$ are the sample sizes of the control and treatment groups, respectively. For simplicity, we let larger outcome values be clinically favorable. We combined two vectors $(x_{1tk}, \ldots x_{n_xtk})$ and $(y_{1tk}, \ldots y_{n_ytk})$, which are the control and treatment groups' observations from outcome $k$ at time $t$, rank them, with larger values obtaining higher ranks, and obtain $R_{xitk}$ and $R_{yjtk}$ being the mid-ranks of observations $x_{itk}$ and $y_{jtk}$, respectively. The midrank is either the regular rank when there is no tie on the observations or the average rank of those tied observations. Further, we define:
\[
\bar{R}_{x\cdot tk}=\frac{1}{n_x}\sum_{i=1}^{n_x}R_{xitk},\quad \bar{R}_{x\cdot t\cdot}=\frac{1}{n_x K}\sum_{i=1}^{n_x}\sum_{k=1}^{K}R_{xitk}, 
\]
and 
\[ \bar{R}_{y\cdot tk}=\frac{1}{n_y}\sum_{j=1}^{n_y}R_{yjtk}, \quad \bar{R}_{y\cdot t\cdot}=\frac{1}{n_y K}\sum_{j=1}^{n_y}\sum_{k=1}^{K}R_{yjtk}, 
\]
for $X$-sample and $Y$-sample, respectively.

To build the LRST procedure, we followed \cite{brunner02book} and define $\theta_{tk}=P(X_{tk}<Y_{tk})-P(X_{tk}>Y_{tk})$ as the relative treatment effect in comparison to the control on outcome $k$ at time $t$, where $X_{tk}$ and $Y_{tk}$ are the random variables corresponding to their observed values $x_{tk}$ and $y_{tk}$. We let $\theta_{t}=\frac{1}{K}\sum_{k=1}^{K}\theta_{tk}$ be the relative treatment effect across all outcomes at time $t$. Notice that $\hat{\theta}_{tk}=\frac{1}{n_x n_y}\sum_{i=1}^{n_x}\sum_{j=1}^{n_y}[I(x_{itk}<y_{jtk})-I(x_{itk}>y_{jtk})]=\frac{2}{N}(\bar{R}_{y\cdot tk}-\bar{R}_{x\cdot tk})$ is a consistent estimator of ${\theta_{tk}}$ (see the detailed derivations in Section S1 of the Supplementary material). It follows that $\hat{\theta}_{t}=\frac{2}{N}(\bar{R}_{y\cdot t\cdot}-\bar{R}_{x\cdot t\cdot})$ is also a consistent estimator of $\theta_{t}$. 

Next, let the rank difference vector $\bm{R}=(\bar{R}_{y\cdot 1\cdot}-\bar{R}_{x\cdot 1\cdot},\cdots, \bar{R}_{y\cdot T\cdot}-\bar{R}_{x\cdot T\cdot})^{\top}$. It is worth noting that $\bm{R}$ is a two-sample U-statistic and thus it follows a multivariate normal distribution asymptotically. In particular,

\allowdisplaybreaks
\begin{align*}
\frac{1}{\sqrt{N}}\left(\begin{array}{c} 
\bar{R}_{y\cdot 1\cdot}-\bar{R}_{x\cdot 1\cdot}  \\
\bar{R}_{y\cdot 2\cdot}-\bar{R}_{x\cdot 2\cdot} \\
\vdots\\
\bar{R}_{y\cdot T\cdot}-\bar{R}_{x\cdot T\cdot}
\end{array}\right)&=\frac{\sqrt{N}}{2n_xn_y}\sum_{i=1}^{n_x}\sum_{j=1}^{n_y}\left(\begin{array}{c} 
\frac{1}{K}\sum_{k=1}^{K}[I(x_{i1k}<y_{j1k})-I(x_{i1k}>y_{j1k})]  \\
\frac{1}{K}\sum_{k=1}^{K}[I(x_{i2k}<y_{j2k})-I(x_{i2k}>y_{j2k})]   \\
\vdots  \\
\frac{1}{K}\sum_{k=1}^{K}[I(x_{iTk}<y_{jTk})-I(x_{iTk}>y_{jTk})]  
\end{array}\right)\\
&=\frac{\sqrt{N}}{2n_xn_y}\sum_{i=1}^{n_x}\sum_{j=1}^{n_y}\bm{h}(\bm{x}_i,\bm{y}_j),
\end{align*}
where $\bm{h}(\bm{x}_i,\bm{y}_j)$ is the kernel function of the above multivariate 2 sample U-statistics and $\bm{x}_i=(x_{i11},\cdots,x_{iTK})^{\top}$, $\bm{y}_j=(y_{j11},\cdots,y_{jTK})^{\top}$,  $\bm{h}_t(\bm{x}_i,\bm{y}_j)=\frac{1}{K}\sum_{k=1}^{K}[I(x_{itk}<y_{jtk})-I(x_{itk}>y_{jtk})]$. By the asymptotic properties of the U-statistics, $\frac{1}{\sqrt{N}}(\bar{R}_{y\cdot 1\cdot}-\bar{R}_{x\cdot 1\cdot},\cdots, \bar{R}_{y\cdot T\cdot}-\bar{R}_{x\cdot T\cdot})^{\top}$ has multivariate normal distribution asymptotically. We established the asymptotic joint distribution of the rank difference vector $\bm{R}$ as follows.

\begin{theorem}
\label{thm1}
The rank difference vector $\frac{1}{\sqrt{N}}\bm{R}$ asymptotically follows a multivariate normal distribution with mean $\frac{\sqrt{N}}{2}(\theta_1,\cdots,\theta_T)^{\top}$ and covariance matrix $\bm{\Sigma}_{T\times T}$ as $\min(n_x,n_y)\rightarrow \infty$ and $\frac{n_x}{n_y}\rightarrow\lambda<\infty$, where $\{\bm{\Sigma}\}_{t_1t_2}$ can be approximated by:
\begin{equation}
\{\bm{\Sigma}\}_{t_1t_2} \approx \frac{1}{K^2}\sum_{k_1=1}^{K}\sum_{k_2=1}^{K}\left[\left(1+\frac{1}{\lambda}\right){c}_{t_1k_1,t_2k_2}+\left(1+{\lambda}\right){d}_{t_1k_1,t_2k_2}\right],  
\label{cov}
\end{equation}
where $c_{t_1k_1,t_2k_2}=cov(G_{t_1k_1}(X_{t_1k_1}),G_{t_2k_2}(X_{t_2k_2}))$ and $d_{t_1k_1,t_2k_2}=cov(F_{t_1k_1}(Y_{t_1k_1}),F_{t_2k_2}(Y_{t_2k_2}))$.
\end{theorem}
Please refer to Section S1 of the Supplementary material for detailed proof. From the expression of $\{\bm{\Sigma}\}_{t_1t_2}$ in Equation~\eqref{cov}, the three sources of correlation for the rank differences ($\bar{R}_{y\cdot tk}-\bar{R}_{x\cdot tk}$) can be explained by the covariance matrix $\bm{\Sigma}$. Specifically, from Equation~\eqref{cov}, 
\[\left(1+\frac{1}{\lambda}\right){c}_{t_1k,t_2k}+\left(1+{\lambda}\right){d}_{t_1k,t_2k},\, k=1,\cdots,K,\, t_1\neq t_2, 
\]
\[\left(1+\frac{1}{\lambda}\right){c}_{tk_1,tk_2}+\left(1+{\lambda}\right){d}_{tk_1,tk_2}, \, t=1,\cdots,T,\, k_1\neq k_2, 
\]
and 
\[\left(1+\frac{1}{\lambda}\right){c}_{t_1k_1,t_2k_2}+\left(1+{\lambda}\right){d}_{t_1k_1,t_2k_2},\, t_1\neq t_2, \, k_1\neq k_2
\]
account for the intra-source correlation (same outcome at different visits), inter-source correlation (different outcomes at the same visit), and cross-correlation (different outcomes at different visits), respectively. Theorem~\ref{thm2} gives the moment estimates of $c_{t_1k_1,t_2k_2}$ and $d_{t_1k_1,t_2k_2}$, and the proof is given in Section S2 of the Supplementary material.

\begin{theorem} 
\label{thm2}
Under the conditions in Theorem~\ref{thm1}, the consistent estimators of $c_{t_1k_1,t_2k_2}$ and $d_{t_1k_1,t_2k_2}$ are given by
\begin{equation}
\hat{c}_{t_1k_1,t_2k_2}=\frac{1}{n_x}\sum_{i=1}^{n_x}\left\{\left[\frac{1}{n_y}\sum_{j=1}^{n_y}I(y_{jt_1k_1}<x_{it_1k_1})-\frac{1-\hat{\theta}_{t_1k_1}}{2}\right]\left[\frac{1}{n_y}\sum_{j=1}^{n_y}I(y_{jt_2k_2}<x_{it_2k_2})-\frac{1-\hat{\theta}_{t_2k_2}}{2}\right]\right\}, 
\label{hat_c}
\end{equation}
and
\begin{equation}
\hat{d}_{t_1k_1,t_2k_2}=\frac{1}{n_y}\sum_{j=1}^{n_y}\left\{\left[\frac{1}{n_x}\sum_{i=1}^{n_x}I(x_{it_1k_1}<y_{jt_1k_1})-\frac{1+\hat{\theta}_{t_1k_1}}{2}\right]\left[\frac{1}{n_x}\sum_{i=1}^{n_x}I(x_{it_2k_2}<y_{jt_2k_2})-\frac{1+\hat{\theta}_{t_2k_2}}{2}\right]\right\}, 
\label{hat_d}
\end{equation}
respectively, where $\hat{\theta}_{tk}=\frac{1}{n_x n_y}\sum_{i=1}^{n_x}\sum_{j=1}^{n_y}[I(x_{itk}<y_{jtk})-I(x_{itk}>y_{jtk})]$, for $t=1,2,\cdots,T$, $k=1,2,\cdots,K$.
\end{theorem}

\subsection{Computation of $\hat{\bm{\Sigma}}$}

We now calculate $\hat{\bm{\Sigma}}$, the estimate of the covariance matrix $\bm{\Sigma}$. For $t_1,t_2\in \{1,2,\cdots,T\}$, from Theorem~\ref{thm1}, the $t_1t_2$th element of $\bm{\Sigma}$ is given by:
\begin{align*}
\{\bm{\Sigma}\}_{t_1t_2}&=cov\left(\frac{1}{\sqrt{N}}(\bar{R}_{y\cdot t_1\cdot}-\bar{R}_{x\cdot t_1\cdot}),\frac{1}{\sqrt{N}}(\bar{R}_{y\cdot t_2\cdot}-\bar{R}_{x\cdot t_2\cdot})\right)\\
& \approx \frac{1}{K^2}\sum_{k_1=1}^{K}\sum_{k_2=1}^{K}\left(\left(1+\frac{1}{\lambda}\right)c_{t_1k_1,t_2k_2}+\left(1+{\lambda}\right)d_{t_1k_1,t_2k_2}\right).\\
\end{align*}
Plugging in the moment estimates $\hat{c}_{t_1k_1,t_2k_2}$ and $\hat{d}_{t_1k_1,t_2k_2}$ in Theorem~\ref{thm2} for $k_1,k_2\in \{1,2,\cdots,K\}$, we obtain the covariance estimate $\hat{\bm{\Sigma}}$.

For ease of computation, let $R_y(x_{itk})$ be the mid-rank of $x_{itk}$ among $\{y_{1tk}, y_{2tk},\cdots,y_{n_ytk}, x_{itk}\}$ and $R_x(y_{jtk})$ be the mid-rank of $y_{jtk}$ among $\{x_{1tk}, x_{2tk},\cdots, x_{n_xtk}, y_{jtk}\}$. Also, we define:
\[
\hat{\theta}_{tk}=\frac{1}{n_x n_y}\sum_{i=1}^{n_x}\sum_{j=1}^{n_y}[I(x_{itk}<y_{jtk})-I(x_{itk}>y_{jtk})],
\]
for $t=1,2,\cdots,T$, $k=1,2,\cdots,K$. Then, $\hat{\theta}_{tk}$ is a consistent estimator of ${\theta}_{tk}$ by the law of large numbers.

Let $\bm{P}_t=\left(P^{(t)}_{ik}\right)_{n_x\times K}$ with $P^{(t)}_{ik}=R_y(x_{itk})-1-n_y(1-\hat{\theta}_{tk})/2$ and $\bm{Q}_t=\left(Q^{(t)}_{jk}\right)_{n_y\times K}$ with $Q^{(t)}_{jk}=R_x(y_{jtk})-1-n_x(1+\hat{\theta}_{tk})/2$, $i=1,2,\cdots, n_x$, $j=1,2,\cdots, n_y$. Then, an empirical estimate of $c_{t_1k_1,t_2k_2}$ is 
\[
\hat{c}_{t_1k_1,t_2k_2}=\frac{1}{n_xn_y^2}\sum_{i=1}^{n_x}P^{(t_1)}_{ik_1}P^{(t_2)}_{ik_2},
\]
which is the expression in Equation~\eqref{hat_c}. Similarly, the estimate of $d_{t_1k_1,t_2k_2}$ is given by:
\[
\hat{d}_{t_1k_1,t_2k_2}=\frac{1}{n_x^2n_y}\sum_{j=1}^{n_y}Q^{(t_1)}_{jk_1}Q^{(t_2)}_{jk_2}, 
\]
which has the expression in Equation~\eqref{hat_d}. Further, define $\bm{C}_{t_1t_2}=(c_{t_1k_1,t_2k_2})_{K\times K}$ and $\bm{D}_{t_1t_2}=(d_{t_1k_1,t_2k_2})_{K\times K}$, $t_1,t_2=1,2,\cdots,T$, $k_1,k_2=1,2,\cdots,K$. Then, it follows that the consistent estimators of $\bm{C}_{t_1t_2}$ and $\bm{D}_{t_1t_2}$ are $\hat{\bm{C}}_{t_1t_2}=\bm{P}^{\top}_{t_1}\bm{P}_{t_2}/(n_x n_y^2)$ and $\hat{\bm{D}}_{t_1t_2}=\bm{Q}^{\top}_{t_1}\bm{Q}_{t_2}/(n_x^2n_y)$, with their elements being $\hat{c}_{t_1k_1,t_2k_2}$ and $\hat{d}_{t_1k_1,t_2k_2}$, respectively. Then, $\{\hat{\bm{\Sigma}}\}_{t_1t_2}$ can be written as:
\[
\{\hat{\bm{\Sigma}}\}_{t_1t_2}=\frac{1}{K^2}\left[\left(1+\frac{1}{\lambda}\right)\cdot sum({\hat{\bm{C}}_{t_1t_2}})+\left(1+{\lambda}\right)\cdot sum({\hat{\bm{D}}_{t_1t_2}})\right],
\]
where $sum({\hat{\bm{C}}_{t_1t_2}})$ and  $sum({\hat{\bm{D}}_{t_1t_2}})$ denote the summation of all the elements of ${\hat{\bm{C}}_{t_1t_2}}$ and ${\hat{\bm{D}}_{t_1t_2}}$, respectively.

\subsection{The longitudinal rank sum test } \label{sec:Methods_LRST}
We let $\bar{\theta}=\frac{1}{T}\sum_{t=1}^{T}{\theta}_t$ be the overall treatment efficacy across all outcomes and time points. If the treatment improves on or slows the progression of some outcomes, we would expect to have $\bar{\theta}>0$. Hence, we tested the overall treatment efficacy with the following null and alternative hypotheses.
\begin{equation}
H_0: \bar{\theta}=0  \quad vs \quad H_1: \bar{\theta}>0.
\label{hypothesis_base}
\end{equation}
For testing the hypothesis in \eqref{hypothesis_base}, we proposed the longitudinal rank-sum test (LRST) statistic
\begin{equation}
T_{LRST}=\frac{\bar{R}_{y\cdot\cdot\cdot}-\bar{R}_{x\cdot\cdot\cdot}}{\sqrt{\widehat{var}(\bar{R}_{y\cdot\cdot\cdot}-\bar{R}_{x\cdot\cdot\cdot})}},
\label{LRST}
\end{equation}
where $\bar{R}_{y\cdot\cdot\cdot}=\frac{1}{n_yTK}\sum_{j=1}^{n_y}\sum_{k=1}^{K}\sum_{t=1}^{T}R_{yjtk}$, $\bar{R}_{x\cdot\cdot\cdot}=\frac{1}{n_xTK}\sum_{i=1}^{n_x}\sum_{k=1}^{K}\sum_{t=1}^{T}R_{xitk}$ are the average rank across all outcomes and time points in the treatment and control groups, respectively. Larger values of the group rank difference $\bar{R}_{y\cdot\cdot\cdot}-\bar{R}_{x\cdot\cdot\cdot}$ suggest treatment efficacy as compared with the control. The denominator, $\widehat{var}(\bar{R}_{y\cdot\cdot\cdot}-\bar{R}_{x\cdot\cdot\cdot})$ is a consistent estimator of the variance of group rank difference $\bar{R}_{y\cdot\cdot\cdot}-\bar{R}_{x\cdot\cdot\cdot}$. The LRST statistic is an advanced rank-sum-type test that enhances existing rank-based tests \citep{OBrien1984Biometrics,Huang2005Biometrics,Liu2010JASA,Zhang2019Biometrics} by effectively leveraging all available longitudinal data.
The non-rejection of the null hypothesis $H_0$, as defined in Equation~\eqref{hypothesis_base}, using the LRST statistic outlined in Equation~\eqref{LRST}, implies the absence of an overall treatment effect across multiple longitudinal outcomes. Conversely, the rejection of $H_0$ suggests that the treatment demonstrates greater effectiveness compared to the control group, providing directional inference. Upon rejection, one may utilize LMMs for each endpoint to identify and quantify treatment effects on specific outcomes.

Next, we derived the asymptotic distribution of the LRST statistic. Let $J$ be a vector of length $T$ with all 1's. From Theorem~\ref{thm1}, $T_{LRST}=\frac{\bar{R}_{y\cdot\cdot\cdot}-\bar{R}_{x\cdot\cdot\cdot}}{\sqrt{{var}(\bar{R}_{y\cdot\cdot\cdot}-\bar{R}_{x\cdot\cdot\cdot})}}=\frac{J^{\top}{\bm{R}}}{\sqrt{J^{\top}{var}(\bm{R})J}}=\frac{J^{\top}{\bm{R}}}{\sqrt{J^{\top}N{\bm{\Sigma}}J}}$ follows a standard normal distribution asymptotically under the null. Incorporating the consistent covariance estimator $\hat{\bm{\Sigma}}$, we established that the proposed longitudinal rank-sum test statistic $T_{LRST}=\frac{\bar{R}_{y\cdot\cdot\cdot}-\bar{R}_{x\cdot\cdot\cdot}}{\sqrt{\widehat{var}(\bar{R}_{y\cdot\cdot\cdot}-\bar{R}_{x\cdot\cdot\cdot})}}
=\frac{J^{\top}{\bm{R}}}{\sqrt{J^{\top}N\hat{\bm{\Sigma}}J}}$ converges in distribution to a standard normal distribution under the null hypothesis, following the conditions specified in Theorem~\ref{thm1}. From the asymptotic normality of $T_{LRST}$, one can evaluate the statistical significance of the proposed test statistic $T_{LRST}$, based on the standard normal distribution. Notice that the test of \cite{Huang2005Biometrics} is a special case of the longitudinal rank-sum test by only using the element $(\bar{R}_{y\cdot T\cdot}-\bar{R}_{x\cdot T\cdot})$ within $\bm{R}$ because it transforms longitudinal data into a cross-sectional format, focusing on changes from baseline to the final observation. Let $J_T=(0,\cdots,0,1)$. Then, substituting $J$ with $J_T$ in $T_{LRST}=\frac{J^{\top}{\bm{R}}}{\sqrt{J^{\top}N\hat{\bm{\Sigma}}J}}$,  the test statistic $T_{LRST}$ reduces exactly to the test statistic of \cite{Huang2005Biometrics}. 

In the simulation and real data analysis, We assessed the performance of the LRST with two other inference methods: (1) the nonparametric factorial ANOVA rank statistics (NFARS) \citep{Brunner2017JRSSB}, implemented using \texttt{R} package \texttt{nparLD} \citep{Noguchi2012JSS_nparLD}, which conducts multiple two-sample univariate tests and generates p-values for each outcome; (2) the univariate LMM, which incorporates covariates including time, treatment, their interaction, with the overall treatment effect determined via the F-test. For both alternative methods, we applied Bonferroni's correction for multiple comparisons, declaring significance when at least one univariate test's p-value fell below the Bonferroni-adjusted significance threshold.

\section{Simulation study} \label{sec:Simulations} 

\subsection{Simulation setting}
Our comprehensive simulation study was designed to rigorously evaluate the LRST test's capability in controlling Type I error rates and its power to detect significant overall treatment effects. The study parameters and structure were modeled after the Bapi 302 clinical trial (as described in Section~\ref{sec:RealData_Bapi302}), incorporating identical primary endpoints (ADAS-cog11 and DAD scores), trial progression patterns, duration, assessment schedules, and a placebo to treatment allocation ratio of 2:3. To simulate the placebo arm data, we first derived from the Bapi 302 study the mean changes from baseline to subsequent visits in ADAS-cog11 and DAD scores, along with their standard deviations (SD), as displayed in Table~\ref{tab:placebo_data_Bapi302}. For each time interval, data were generated using a piecewise linear mixed model with the mean changes and SDs in Table~\ref{tab:placebo_data_Bapi302}. We modeled within-subject temporal variability by integrating random effects with standard deviations matching those observed in the study, employing an AR(1) structure with a lag-1 autocorrelation coefficient of 0.6. 

\begin{table}[htbp] \footnotesize
    \centering
    \begin{tabular}{c c c c c c c } \hline
        Visit Week & 13   & 26   & 39   & 52   & 65   & 78   \\ \hline
        ADAS-cog11 & $0.739_{4.799}$ & $1.322_{5.386}$ & $3.166_{6.510}$ & $4.607_{7.444}$ & $5.899_{8.084}$ & $7.457_{9.139}$ \\ 
        DAD        & $-0.706_{10.561}$ & $-4.065_{13.057}$ & $-5.705_{14.960}$ & $-8.249_{15.662}$ & $-12.104_{16.940}$ & $-13.941_{18.080}$ \\ \hline
    \end{tabular}
    \caption{Mean changes from baseline to each visit for ADAS-cog11 and DAD scores, with standard deviations in subscripts. Abbreviation: ADAS-cog11, Alzheimer’s Disease Assessment Scale - cognitive subscale (11 items); DAD, Disability Assessment for Dementia. }
    \label{tab:placebo_data_Bapi302}
\end{table}

Furthermore, recognizing the LRST's capacity to analyze both continuous and ordinal outcomes, our study extends its application to ordinal data, which often presents extensive ties. To investigate the test's robustness under such conditions, we transformed simulated continuous data into ordinal categories (0 through 4). Specifically, continuous simulated data were discretized into ordinal categories (0, 1, 2, 3, 4), utilizing thresholds set at $\mu - 3 \sigma, \mu - \sigma, \mu+ \sigma, \mu+ 3 \sigma$, where $\mu$ and $\sigma$ reference the means and standard deviations from Table~\ref{tab:placebo_data_Bapi302}. The application of LRST to ordinal data derived from continuous simulations is denoted as LRST-ordinal. This extension allows us to assess the LRST's adaptability and performance in handling the complexities associated with multiple ordinal longitudinal outcomes in clinical research settings. 

\subsection{Simulation results} \label{sec:Simulations_results} 

\subsubsection{Type I error} \label{sec:Simulations_results_TypeI} 
To assess Type I error, we conducted simulations assuming no treatment effects across both outcomes and throughout all visits for both groups. To introduce inter-source correlation between the two outcomes, we set the between-outcome correlation coefficient as 0.5. The empirical Type I error rates of the LRST, at nominal levels of $\alpha=0.05$ and $0.1$, across various sample sizes ($N$) from 1000 simulations, are presented in Table~\ref{tab:typeI}. These results closely align with the specified $\alpha$ levels, demonstrating the LRST's capacity to maintain accurate Type I error rates, even in scenarios with smaller sample sizes, e.g., $N=100$. Moreover, when the continuous simulated data were converted into ordinal outcomes, the LRST effectively controlled Type I error, underscoring its adeptness at managing the complexities introduced by significant ties in multivariate longitudinal ordinal data.

\begin{table}[htbp] \centering
\begin{tabular}{ccccc} \hline
 & \multicolumn{2}{c}{LRST} & \multicolumn{2}{c}{LRST-ordinal} \\ \cline{2-5} 
N & $\alpha$ = 0.05  & $\alpha$ = 0.1 & $\alpha$ = 0.05 & $\alpha$ = 0.1 \\ \hline
100  & 0.058 & 0.104 & 0.055 & 0.096 \\
300  & 0.047 & 0.099 & 0.050 & 0.106 \\
500  & 0.049 & 0.100 & 0.054 & 0.109 \\
700  & 0.053 & 0.093 & 0.054 & 0.107 \\
900  & 0.047 & 0.092 & 0.051 & 0.099 \\
1200 & 0.059 & 0.106 & 0.054 & 0.097 \\
1500 & 0.530 & 0.101 & 0.051 & 0.094 \\ \hline
\end{tabular}
\caption{Comparison of empirical Type I error rates between LRST for continuous data and LRST-ordinal for discretized data across various sample sizes ($N$) and nominal levels of $\alpha$. The placebo-to-treatment allocation ratio was 2:3. The results are based on 1,000 simulations.}
\label{tab:typeI}
\end{table}

\subsubsection{Power} \label{sec:Simulations_results_Power}
In our power analysis, we considered two scenarios. In Scenario 1, we assumed that the Bapi group would exhibit the pre-specified effect sizes (2.21 unit advantage in ADAS-cog11 and 5.38 unit advantage in DAD over placebo, equally distributed across each visit) and set the between-outcome correlation coefficient as 0.5. We investigated the power under various total sample sizes ($N=100, 300, 500, 700, 900, 1200, 1500$) with a fixed allocation ratio of 2:3 to the placebo and treatment groups. In Scenario 2, we fixed the total sample size at $N=759$ (mirroring the actual sample sizes of the placebo ($n_x=311$) and Bapi ($n_y=448$) groups in our final analysis dataset). We varied the effect size to range from 0.2 to 2 times that specified in Scenario 1 (denoted as ``0.2 to 2 deviations'') and set the between-outcome correlation coefficient to be 0, 0.2, 0.5, and 0.8. In both scenarios, we conducted 1,000 simulations of the RCT. 

\begin{figure}[htbp]
    \centering
    \includegraphics[width = \textwidth]{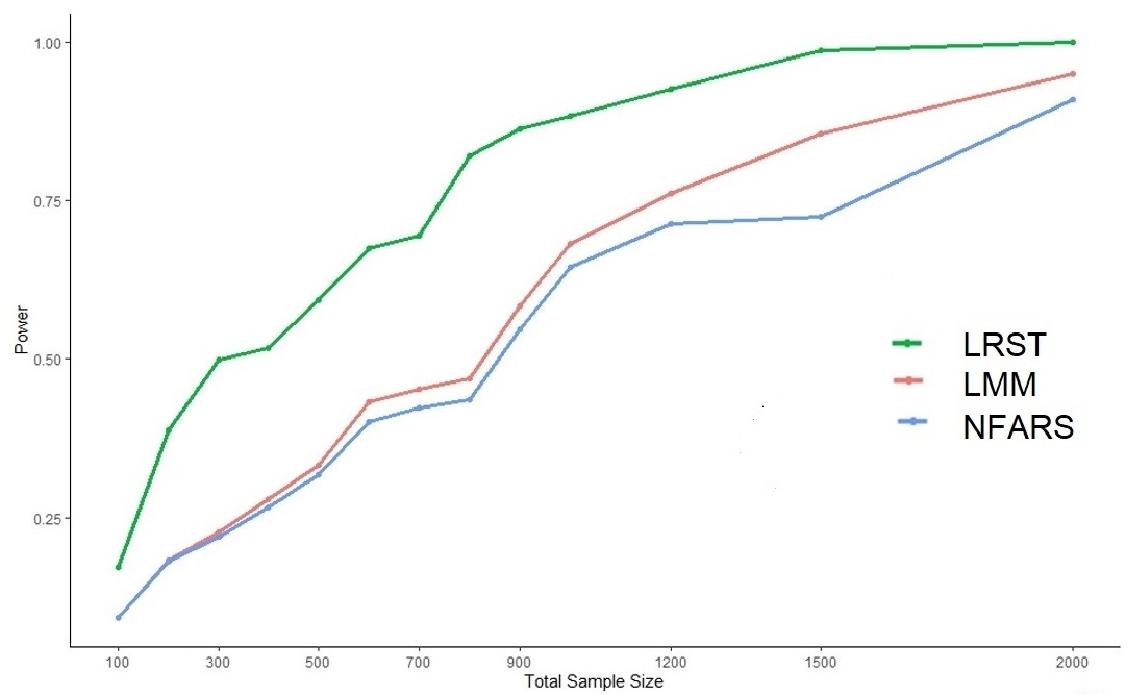}
    \caption{Empirical power curves of LRST, NFARS, and LMM methods for varying total sample sizes $N$ in Scenario 1. The placebo-to-treatment allocation ratio was 2:3. Effect sizes were preset at 2.21 for ADAS-cog11 and 5.38 for DAD. Abbreviation: NFARS, the nonparametric factorial ANOVA rank statistics \citep{Brunner2017JRSSB}; LMM, linear mixed model; ADAS-cog11, Alzheimer’s Disease Assessment Scale - cognitive subscale (11 items); DAD, Disability Assessment for Dementia. } \label{fig:Sample-Size}
\end{figure}

Figure~\ref{fig:Sample-Size} displays the empirical power curves of the LRST, NFARS, and LMM methods across different total sample sizes in Scenario 1. The proposed LRST test consistently has higher power than the NFARS and LMM tests across all sample sizes considered. For example, at $N = 900$, LRST reached a power of 0.863, significantly surpassing the LMM and NFARS with powers of 0.584 and 0.548, respectively. This result highlights the efficacy of the LRST in evaluating overall treatment impact across multiple longitudinal outcomes, proving to be more powerful than individual univariate parametric or nonparametric tests that require adjustments for multiple comparisons. Notably, the LMM method exhibited marginally greater power compared to NFARS across most sample sizes, due to the simulation data being generated under the piecewise LMM framework. Table~\ref{tab:power_ordinal} provides a comparison of empirical power between LRST applied to continuous data and LRST on ordinal data discretized from the same continuous simulations in Scenario 1. Despite the introduction of extensive ties through discretization, the LRST applied to ordinal data exhibits slightly lower yet remarkably close power to that of its continuous counterpart, confirming the LRST's robust performance in scenarios characterized by significant ties in multivariate longitudinal ordinal outcomes.

\begin{figure}[htbp]
    \centering
    \includegraphics[width = \textwidth]{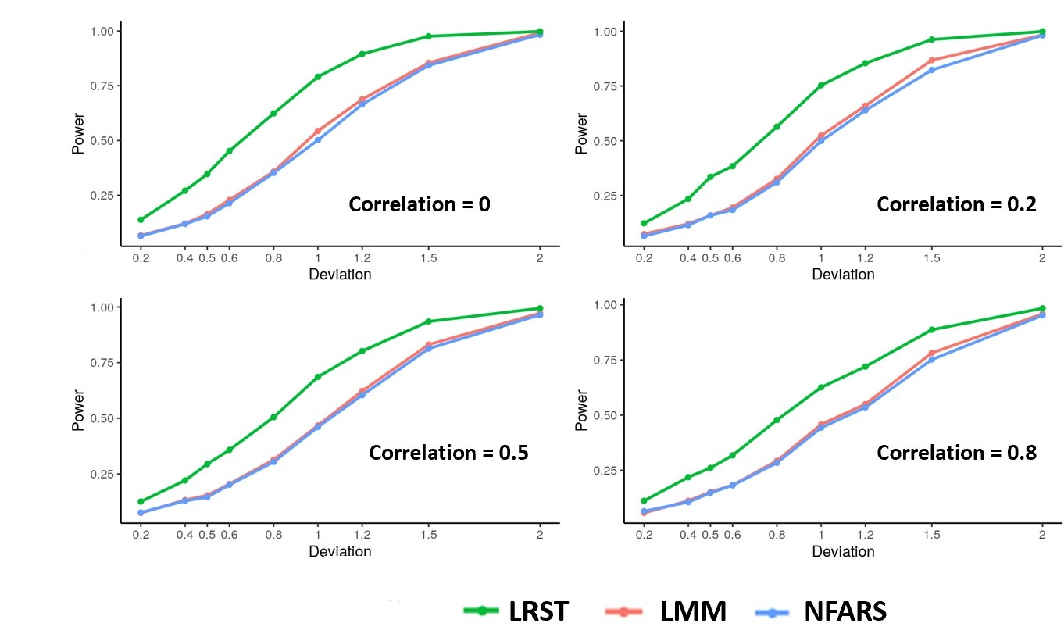}
    \caption{Empirical power curves of LRST, NFARS, and LMM methods under Scenario 2. This figure illustrates the power profiles for varying effect sizes, i.e., from 0.2 to 2 times those in Scenario 1 (labeled as ``0.2 to 2 deviations''), across different inter-outcome correlation coefficients (0, 0.2, 0.5, and 0.8). The sample size is $N=759$ with $n_x=311$ in the placebo group and $n_y=448$ in the Bapi group. Abbreviation: NFARS, the nonparametric factorial ANOVA rank statistics \citep{Brunner2017JRSSB}; LMM, linear mixed model.}
    \label{fig:PowerCurveCorrelation}
\end{figure}

Figure~\ref{fig:PowerCurveCorrelation} presents the empirical power curves of the LRST, NFARS, and LMM methods under various effect sizes and inter-outcome correlations in Scenario 2. Throughout these configurations, the LRST consistently demonstrates significantly greater power than both LMM and NFARS, underscoring its superior efficacy in detecting treatment effects across various sample sizes and correlations. This confirms LRST's advantage in evaluating treatment effects across multiple longitudinal outcomes, surpassing the capabilities of univariate tests adjusted for multiple comparisons. LMM's marginally higher power relative to NFARS is again due to the simulation framework favoring the LMM approach.

\begin{table}[htbp]
    \centering
    \begin{tabular}{c c c c c c c c} \hline
        N            & 100   & 300   & 500   & 700   & 900   & 1200  & 1500 \\ \hline
        LRST         & 0.171 & 0.499 & 0.593 & 0.726 & 0.863 & 0.926 & 0.987 \\ 
        LRST-ordinal & 0.172 & 0.481 & 0.584 & 0.696 & 0.823 & 0.893 & 0.929 \\ \hline
    \end{tabular}
    \caption{Comparison of empirical power between LRST for continuous data and LRST-ordinal for discretized data across various sample sizes ($N$) in Scenario 1. The placebo-to-treatment allocation ratio was 2:3. Effect sizes were preset at 2.21 for ADAS-cog11 and 5.38 for DAD. Abbreviation: ADAS-cog11, Alzheimer’s Disease Assessment Scale - cognitive subscale (11 items); DAD, Disability Assessment for Dementia. }
    \label{tab:power_ordinal}
\end{table}

\section{Real Data Application in Neurodegenerative Disorders} \label{sec:RealData}

\subsection{Analysis of the Bapineuzumab 302 trial} \label{sec:RealData_Bapi302}
The Bapineuzumab (Bapi) 302 study \citep{Salloway2014NEJM_bapineuzumab} is a double-blind, randomized, placebo-controlled, phase 3 RCT aimed at evaluating the efficacy of intravenous Bapi as compared with placebo in patients with mild-to-moderate AD dementia. Conducted at 170 sites in the United States from December 2007 through April 2012, the study employed a 2:3 allocation ratio to randomize participants into placebo and Bapi treatment groups. Assessments were conducted at baseline, subsequently every 13 weeks over 78 weeks ($t=13, 26, 39, 52, 65, 78$ weeks with $T=6$). 

The primary endpoints of the study were the scores on the 11-item cognitive subscale of the Alzheimer’s Disease Assessment Scale (ADAS-cog11) and the Disability Assessment for Dementia (DAD). The ADAS-cog11, a comprehensive tool consisting of 11 items, is designed to evaluate cognitive functions in AD patients across various domains including memory, language, praxis, and orientation. This is accomplished through a series of tasks ranging from word recall to object naming and command following. Scores on the ADAS-cog11 vary from 0 to 70, where higher scores denote more severe cognitive impairment, thus aiding clinicians in diagnosis and monitoring of disease progression. On the other hand, the DAD score, derived from caregiver assessments over 40 items, quantifies the patient's autonomy in daily tasks, such as personal hygiene and meal preparation. This scale ranges from 0 to 100, where a score of 100 reflects complete independence in all assessed activities, providing a metric to evaluate functional abilities and tailor interventions accordingly. The study's dataset comprised $759$ participants, with $311$ ($n_x$) in the placebo arm and $448$ ($n_y$) receiving Bapi treatment. 

\begin{figure}[htbp]
    \centering
    \includegraphics[width = \textwidth]{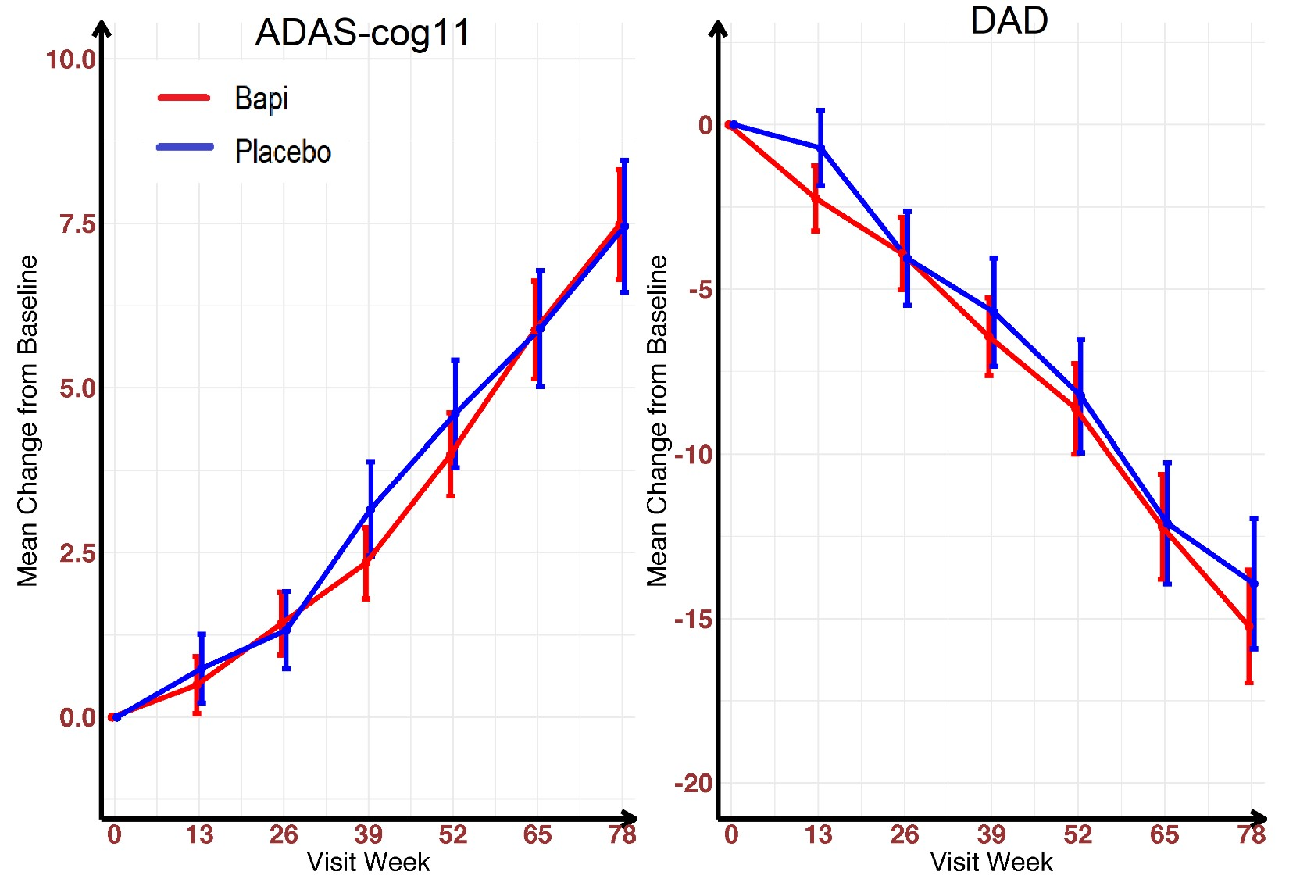}
    \caption{Mean changes from baseline to week 78 in scores on ADAS-cog11 (left panel, with scores ranging from 0 to 70 and higher scores indicating greater impairment) and DAD (right panel, with scores ranging from 0 to 100 and higher scores indicating less impairment) in the Bapineuzumab 302 Trial. Vertical lines are 95\% confidence intervals. Abbreviation: ADAS-cog11, Alzheimer’s Disease Assessment Scale - cognitive subscale (11 items); DAD, Disability Assessment for Dementia.}
    \label{fig:mean-curve}
\end{figure}

Figure~\ref{fig:mean-curve} displays the mean changes from baseline for both ADAS-cog11 and DAD scores in the placebo and Bapi treatment groups. The increase in ADAS-cog11 scores over 78 weeks suggests an ongoing decline in cognitive function, while the concurrent decrease in DAD scores indicates a continuous loss of daily functioning capabilities in participants. These trends, indicative of the gradual deterioration characteristic of AD in both cognitive and functional domains, are observable in both treatment groups. However, no significant differences in the primary outcomes between the Bapi and placebo groups were observed. The LRST statistic was $-0.0474$ (SE: $1.950$), resulting in a Z-statistics of $-$0.024 ($P=0.509$), suggesting no significant differences in scores between the Bapi and the placebo groups across cognitive and functional domains. 

In comparison, the nonparametric NFARS method gave p-values of $0.500$ (ADAS-cog11) and $0.427$ (DAD). As the minimum of these p-values (0.427) exceeds the Bonferroni-adjusted significance level of 0.05/2=0.025, we do not reject the null hypothesis of no treatment efficacy. Moreover, using the LMM method, the p-values derived from the F-test for the Bapi effect in the mixed model are 0.740 (ADAS-cog11) and 0.473 (DAD). After adjusting for multiple testing, none of these p-values are less than 0.05/2=0.025, leading to a failure to reject the null hypothesis of no treatment effect. Hence, all three analytical approaches yield results consistent with the findings of the Bapi RCT, indicating that Bapi did not improve clinical outcomes related to cognition and daily function in patients with AD \citep{Salloway2014NEJM_bapineuzumab}. 

\subsection{Analysis of the Azilect Study} \label{sec:RealData_Azilect}
The Azilect study \citep{Hattori2019PRD} is a Phase 3, randomized, double-blind, placebo-controlled clinical trial designed to evaluate the efficacy of rasagiline in Japanese patients with early-stage Parkinson's Disease (PD), the second most prevalent neurodegenerative disorder. This trial enrolled patients aged between 30 and 79 years, diagnosed with PD within the previous five years. Participants were assigned randomly in a 1:1 ratio to receive either 1 mg/day of rasagiline or a placebo for up to 26 weeks. Assessments occurred at baseline, followed by weeks 6, 10, 14, 20, and 26 ($T=5$). The primary endpoint was the Movement Disorder Society Unified Parkinson's Disease Rating Scale (MDS-UPDRS), a widely recognized scale for assessing Parkinsonian symptoms in both clinical and research contexts \citep{Regnault2019JN}. The MDS-UPDRS includes 65 items across four sections, using a 5-point Likert scale (0-4, with higher scores indicating greater severity): Part I, Non-Motor Aspects of Experiences of Daily Living (13 items); Part II, Motor Aspects of Experiences of Daily Living (13 items); Part III, Motor Examination (33 items); and Part IV, Motor Complications (6 items). Additional details on the MDS-UPDRS scale are available in \citep{Goetz2008MDS}.

\begin{figure}
\centering
\includegraphics[width=\textwidth]{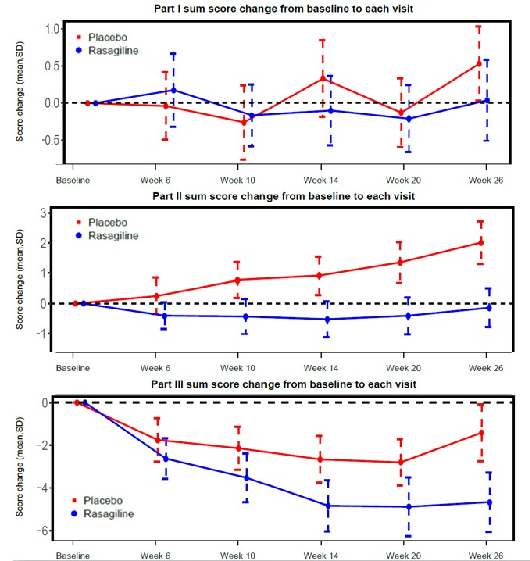}
\caption{Mean changes from baseline to week 26 in MDS-UPDRS Parts I, II, and III sum scores. The vertical lines are the 95\% confidence intervals. The dashed horizontal line indicates the baseline score reference. Dotted and dashed vertical lines represent standard deviations. Abbreviations: SD, standard deviation; MDS-UPDRS, Movement Disorder Society Unified Parkinson's Disease Rating Scale. \label{score}}
\end{figure}

We applied our proposed statistical tests to the multiple outcomes derived from the MDS-UPDRS scale, focusing on the sum scores for Parts I (0 to 52), II (0 to 52), and III (0 to 132), thereby considering three primary outcomes ($K=3$). Out of 209 patients ($N=209$) who completed all study visits, 100 were in the placebo group ($n_x=100$) and 109 received rasagiline ($n_y=109$). Figure~\ref{score} illustrates the mean changes from baseline to week 26 in MDS-UPDRS Parts I, II, and III sum scores. Notably, the mean change in the MDS-UPDRS Part I sum score from baseline was similar across both groups. However, while the placebo group showed a consistent increase in the MDS-UPDRS Part II sum score change from baseline (indicating deterioration), the rasagiline group exhibited some improvements during the study, as compared to the baseline, although the improvements diminished towards the study's conclusion. The treatment effect of rasagiline on the MDS-UPDRS Part II sum score demonstrated a growing trend over the study period compared to placebo. For the MDS-UPDRS Part III sum score change from baseline, both groups showed steady improvement over the study duration, with the rasagiline group exhibiting more significant improvement and efficacy than the placebo group.

To align higher scores with clinically better outcomes, as defined in Section~\ref{sec:Methods}, we inverted all ordinal item scores (multiplied by $-1$) to calculate sum scores. The LRST statistic was $4.929$ (SE: $1.572$), resulting in a Z-statistic of $3.135$ ($P=8.592 \times 10^{-4}$), indicating a significant overall treatment effect of rasagiline on the primary outcomes of MDS-UPDRS Parts I, II, and III compared to placebo. Following the significant overall treatment effect, we fitted three univariate LMMs using MDS-UPDRS Parts I, II, and III sum scores as response variables. The p-values obtained from the F-test for the rasagiline effect in the mixed model are $0.418$ (Part I), $2.95 \times 10^{-4}$ (Part II), and $2.03 \times 10^{-5}$ (Part III). Since the last two p-values are below the Bonferroni-adjusted significance level of 0.05/3=0.017, we concluded that rasagiline significantly improved motor aspects of daily living and motor examination in MDS-UPDRS Parts II and III, but did not effectively improve non-motor aspects of daily living in Part I.

Additionally, we applied nonparametric NFARS method to MDS-UPDRS Parts I, II, and III sum scores separately. The resulting p-values for these tests are $0.174$, $0.0001$, and $0.0013$. Given that only the p-values for MDS-UPDRS Parts II and III sum scores were below the Bonferroni-adjusted significance level of 0.05/3=0.017, our conclusions are consistent with those from the LRST. Both the LRST and nonparametric NFARS corroborate the visible group differences in Parts II and III score changes (Figure~\ref{score}, middle and lower panels), with minimal differences in Part I score changes (Figure~\ref{score}, upper panel).

\section{Discussions} \label{sec:Discussions}

This article develops and validates the longitudinal Rank Sum Test (LRST), a novel nonparametric rank-based hypothesis-testing statistic, for analyzing multiple longitudinal primary endpoints in randomized controlled trials of neurodegenerative disorders such as AD and PD. The LRST provides a robust and efficient statistical framework for assessing treatment efficacy and capturing dynamic changes in multiple endpoints over time. Our simulation study revealed that the LRST significantly outperforms traditional univariate tests requiring multiple comparison adjustments, such as Linear Mixed Models (LMM) and nonparametric factorial ANOVA rank statistics, by achieving higher statistical power in identifying overall treatment effects while precisely controlling Type I error rates. Applied to data from the Bapineuzumab AD study and the Azilect PD study, the LRST effectively detected overall treatment benefits, aligning with the primary conclusions of these studies. Following the determination of overall treatment efficacy, methodologies like LMM or Generalized Estimating Equations (GEE) can be employed to identify and quantify the treatment effects on individual outcomes.

The LRST statistic assigns equal weight across all time points and outcomes. To enhance the analysis's clinical relevance, we introduce a weighting mechanism that assigns variable weights over time, e.g., larger weights to in-clinic over telemedicine visits. To facilitate this, we may define a weighted sum of rank differences between groups across time points, denoted by $\bm{R}^{\bm{w}}$=$(w_1(\bar{R}_{y\cdot 1\cdot}-\bar{R}_{x\cdot 1\cdot}),\cdots, w_T(\bar{R}_{y\cdot T\cdot}-\bar{R}_{x\cdot T\cdot}))^{'}$, where the weight $w_t\geq 0$ and $\sum_{t=1}^{T}w_t=1$. We then follow the procedure in Section~\ref{sec:Methods} to develop a weighted LRST. This weighting methodology allows the LRST to more accurately capture the clinical significance of each visit type and can be applied to various outcomes. The weighted LRST simplifies to the original LRST when weights are equal. Moreover, optimizing weights via the covariance estimator $\hat{\bm{\Sigma}}$ may lower the weighted LRST's asymptotic variance, consistent with previous findings \citep{Wei1985Biometrika, Wei1989JASA}. Development of the weighted LRST represents a future research direction.

One major challenge in the longitudinal studies is the issue of missing data. Under missing at random (MAR) and missing not at random (MNAR) cases, ignoring the missing data may lead to biased estimates and invalid inference. To this end, multiple imputation and inverse probability weighting techniques can be used to handle missing data. By using appropriate multiple imputation techniques and models, multiple complete datasets are generated so that the LRST can be applied. Under the weighting paradigm, the kernel function $\bm{h}(\bm{x}_i,\bm{y}_j)$ (defined in Section S1 in the Supplementary material) of the U-statistic needs to be weighted by the estimated inverse probabilities to account for the missing outcome data. The development of consistent estimators for response probabilities and their integration into the longitudinal rank-sum test procedure remains an area for future research.

\section{Software}

The R package of LRST and implementation codes are available at: \href{https://github.com/djghosh1123/LRST/tree/master}{https://github.com/djghosh1123/LRST}.

\section*{Acknowledgement}
The research of Sheng Luo was supported by National Institute on Aging (grant numbers: R01AG064803 and P30AG072958). 

This study, carried out under YODA Project \#2020-4448, used data of the Bapineuzumab 302 study obtained from the Yale University Open Data Access Project, which has an agreement with JANSSEN RESEARCH \& DEVELOPMENT, L.L.C.. The interpretation and reporting of research using this data are solely the responsibility of the authors and do not necessarily represent the official views of the Yale University Open Data Access Project or JANSSEN RESEARCH \& DEVELOPMENT, L.L.C..

The Azilect study data are available from the Critical Path Institute’s CPP Consortium. The Critical Path Institute’s CPP Consortium is funded by Parkinson’s United Kingdom and the following industry members: AbbVie, Biogen, Cerevel, Denali, GSK, MSD, Takeda, Sanofi, Roche, IXICO, Cereval, Clario and UCB. We also acknowledge additional CPP member organizations, including the Parkinson’s Disease Foundation, The Michael J. Fox Foundation, the Davis Phinney Foundation, The Cure Parkinson’s Trust, PMD Alliance, the University of Oxford, University of Cambridge, Newcastle University, University of Glasgow, as well as the NINDS, US Food and Drug Administration, and the European Medicines Agency. 

\section*{Declaration of conflicting interests}
The author(s) declared no potential conflicts of interest with respect to the research, authorship, and/or publication of this article. 

\bibliography{finalRef}
\bibliographystyle{rusnat}


\end{document}